\documentclass[%
 reprint,
 amsmath,amssymb,
 aps,
]{revtex4-2}

\usepackage{graphicx}
\usepackage{dcolumn}
\usepackage{bm}
\usepackage{url}
\usepackage{hyperref}
\usepackage[dvipsnames]{xcolor}
\definecolor{blue(pigment)}{rgb}{0.2, 0.2, 0.6}
\usepackage{bbold}
\hypersetup{
    allcolors=blue(pigment),
    colorlinks=true,
    linkcolor=blue(pigment),
    filecolor=blue(pigment),      
    urlcolor=blue(pigment),
}

\newcommand{\Rnum}[1]{\lowercase\expandafter{\romannumeral #1\relax}}
\newcommand{\RNum}[1]{\uppercase\expandafter{\romannumeral #1\relax}}

\begin{document}
\begin{flushright}
{KOBE-COSMO-26-05}
\end{flushright}
\preprint{APS/123-QED}

\title{Cavity-QED Transducer of Gravitons
}


\author{F. Shojaei Arani}
\affiliation{Department of Physics, University of Isfahan, Hezar Jerib Str., Isfahan 81746-73441, Iran}

\author{Brahim Lamine}
\affiliation{Université de Toulouse, OMP, IRAP, F-31400 Toulouse, France}

\author{Jiro Soda}
\affiliation{Department of Physics, Kobe University, Kobe 657-8501, Japan}





\begin{abstract}
We develop a quantum description of the resonant interaction between electromagnetic (EM) and gravitational waves (GW).
We first show that Lorentz invariance together with polarization selection rules forbids any photon-graviton mixing in free space.
We demonstrate that confining the EM field within a cavity quantum electrodynamics (cavity-QED) environment breaks translational symmetry and isotropy, leading to non-vanishing mode coupling between EM and gravitational degrees of freedom. 
Within this framework, we identify multiple photon-graviton scattering channels, including photon up- and down-conversion and photon creation.
In the semiclassical limit of the trilinear interaction where GW acts as a classical pump and the EM field is in a vacuum, spontaneous parametric photon amplification and two-mode squeezing occur. When the gravitational field is quantized, however, the back-action and energy exchange between photons and gravitons result in saturation of amplification, in contrast to exponential growth, and the loss of purity in the photonic subsystem. The characteristic timescale scales as $t_{\text{sp}}\sim (g\sqrt{n_g})^{-1}$, where $g$ and $n_g$ refer to the coupling strength and the mean graviton number, demonstrating collective enhancement of the interaction with the graviton occupation number. In the stimulated regime, where one EM mode is initially populated, the effective coupling is further enhanced, analogous to Dicke-type superradiant emission. 
This work introduces a cavity-based graviton transducer for probing quantum aspects of GWs.
\end{abstract}

\keywords{Parametric resonance, gravitational waves, photon-graviton exchange}
\maketitle



\section{\label{sec:1}Introduction}

The question of whether gravity must be quantized has long been debated through both theoretical and experimental arguments (see, e.g., \cite{dewitt1967quantum,eppley1977necessity,page1981indirect,carlip2008quantum,stamp2015rationale}). In a seminal work, Page and Geilker \cite{page1981indirect} demonstrated that a semiclassical model in which a classical gravitational field couples to the expectation value of the quantum energy-momentum tensor of matter is inconsistent with observed macroscopic outcomes. While this result does not constitute a definitive proof of gravity quantization, it highlights fundamental tensions between classical gravity and quantum matter.
On the other hand, Freeman Dyson \cite{dyson2012graviton} addressed the complementary question of detectability: even if gravitons exist and gravity is fundamentally quantized, can individual quanta of the gravitational field ever be observed? Dyson examined a range of conceivable detectors (i) interferometric systems like LIGO, (ii) atomic-scale absorbers, and (iii) mechanisms based on photon-graviton mixing in the presence of large-scale magnetic fields, first proposed by Gertsenshtein \cite{Gertsenshtein1962}. Dyson then concludes that detecting gravitons using a LIGO-like interferometer is ruled out because such a detector with the required sensitivity would inevitably collapse into a black hole.
Other methods based on atoms are incapable of graviton detection due to the background noise in atomic detectors. The third class of detectors considered by Dyson based on the Gertsenshtein mechanism does not work due to the QED effect for a strong magnetic field.
Consequently, according to Dyson, even if the gravitational field is quantized, direct detection of individual gravitons remains fundamentally challenging.

These considerations motivate the search for alternative mechanisms that enable photon-graviton coupling without relying on strong external magnetic fields. In this work, we explore such a mechanism by considering a cavity quantum electrodynamic (cavity-QED) environment, in which conducting boundaries provide the necessary symmetry breaking to mediate photon-graviton coupling.

Several proposals have investigated laboratory and quantum-optical schemes for detecting GWs based on electromagnetic transduction, often in the context of the inverse Gertsenshtein effect. It has been suggested that GWs could generate a weak photon flux in the presence of strong background magnetic fields, potentially observable in microwave or optical cavities \cite{Pegoraro1978,Li2008}, in particular, in the case of stimulated conversion~\cite{Ikeda:2025uae}. Related resonant detector concepts have been explored in the MHz-GHz frequency range, where conventional interferometers lose sensitivity \cite{Cruise2006}. More recently, analogous ideas have been developed using cavity optomechanics, superconducting circuits, and other hybrid quantum systems \cite{Baker2003,Rueda2019,kanno2025search,Aggarwal2021}. Together, these works illustrate a broad theoretical interest in EM systems as transducers of gravitational signals and motivate the development of new conceptual frameworks for EM-GW coupling beyond large external magnetic fields.

A particularly intriguing aspect of EM-GW interaction is the possibility of parametric resonance, i.e., the amplification of oscillations of a system driven at multipliers of its natural frequencies \cite{Mathieu1868, Landau1976}. Parametric amplification is central to quantum optics, underlying squeezed-state generation and three-wave mixing (3WM)\cite{Loudon2000,WallsMilburn2008}, and also plays a key role in the dynamical Casimir effect, where vacuum fluctuations are converted into real photons \cite{Moore1970,lambrecht1996motion,Dodonov2010,Wilson2011}. By analogy, it is expected that GWs themselves may act as a parametric drive for EM fields. This idea was explored recently in Ref.~\cite{brandenberger2023graviton}, where a Mathieu-type equation governing classical EM dynamics leads to exponential amplification under resonant conditions. Still, a fully quantum description based on a trilinear interaction Hamiltonian, which makes parametric-resonance and 3WM dynamics explicit and allows a systematic comparison between semiclassical amplification and fully quantum behavior, has been absent. In particular, the role of back-action, saturation, and quantum correlations in quantum parametric resonance has not been explored in EM-GW interaction. Specially, a central consequence of the trilinear interaction is the role of pump depletion. In the standard parametric approximation, the gravitational pump is assumed undepleted, leading to unbounded exponential growth of the EM modes \cite{brandenberger2023graviton}. However, when the gravitational field is treated quantum mechanically, energy exchange between photons and gravitons becomes unavoidable. As a result, exponential amplification is replaced by oscillatory conversion dynamics and eventual saturation, reflecting the depletion of the gravitational pump. 
The present work addresses this gap by formulating the cavity QED-GW system in Hamiltonian language and identifying the conditions in which parametric photon generation and mode mixing can occur. 

An alternative route has been recently proposed based on analyzing the possibility of stimulated emission and absorption of gravitons by light in interferometric geometries, like LIGO \cite{schutzhold2025stimulated}. In that approach, a GW induces tiny frequency shifts in propagating EM fields, which may be interpreted as stimulated graviton exchange when the optical field is treated quantum mechanically. The effect accumulates coherently over extended optical paths and relies on precise phase control and interference. 
The cavity-QED mechanism developed here is conceptually distinct: instead of relying on phase accumulation along extended paths, confinement of the EM field within a cavity introduces discrete resonant modes and a well-defined interaction time. This enables a resonant 3WM process between two EM modes and one gravitational mode, governed by a trilinear Hamiltonian. In this setting, the interaction is not merely a perturbative frequency shift but a genuine energy-exchange process. While both approaches exploit stimulated processes, the cavity-based framework naturally exposes the nonlinear and fully quantum dynamics of photon-graviton coupling, providing access to regimes that cannot be captured by free-space or interferometric descriptions.


\section{\label{sec:2}Prohibition of photon-graviton mixing in free space}

GWs are perturbations to the background spacetime metric described by $g_{\mu\nu}=\eta_{\mu\nu}+h_{\mu\nu}$. The quantization of GWs is done in the traceless-transverse (TT) gauge and the plane wave expansion of a mono-mode GW with wave vector $\mathbf{K}$ and polarization state $\gamma$, in terms of the ladder operators $\hat{b}_{\mathbf{K},\gamma}, \hat{b}^{\dagger}_{\mathbf{K},\gamma}$ is
\begin{eqnarray}\label{eq:1}
\hat{h}_{ij}^{TT}(\mathbf{x},t) &=& \mathcal{C}_g\, e_{ij}^{\gamma}(\mathbf{K}) \big( \hat{b}_{\mathbf{K},\gamma} e^{-i(\Omega t -\mathbf{K} \cdot \mathbf{x})} + \hat{b}^{\dagger}_{\mathbf{K},\gamma} e^{i(\Omega t -\mathbf{K} \cdot \mathbf{x})} \big),\nonumber\\
\end{eqnarray} 
where $\mathcal{C}_g \equiv \frac{\hbar}{E_{pl}} \sqrt{\frac{16 \pi c^3}{2\Omega_K V}}$, and the polarization tensor $e_{ij}^{\gamma}(\mathbf{K})$ is transverse-traceless. The TT gauge description is useful when considering free space propagation of EM fields or assuming freely-falling detectors.
Within the linearized Einstein theory, the interaction between EM and GW fields is described by the energy-momentum tensor of the EM field, $T^{\mu\nu}$, through the action $S_{\text{int}} = \frac{(32\pi G)^{1/2}}{2} \int d^4x \, h_{\mu\nu}T^{\mu\nu}$. Following the approach introduced in ~\cite{pang2018quantum,arani2025revisiting,arani2025phase}, it turns out that the interaction Hamiltonian takes the general form
\begin{eqnarray}\label{eq:2}
\hspace*{-0.2cm}\hat{H}_{int} / \hbar&\supset& \sum_{k,k',K} \bigg\{ \mathcal{G}_{\text{up}}(k,k',K)\, \hat{a}_{k} \hat{a}^{\dagger}_{k'} \hat{b}_{K} \nonumber\\
&+& \mathcal{G}_{\text{down}}(k,k',K)\, \hat{a}_{k} \hat{a}^{\dagger}_{k'} \hat{b}^{\dagger}_{K} \nonumber \\
&-& \mathcal{G}_{\text{mix}}(k,k',K)\,\Big( \hat{a}_{k} \hat{a}_{k'} \hat{b}^{\dagger}_{K} + \hat{a}^{\dagger}_{k} \hat{a}^{\dagger}_{k'} \hat{b}_{K} \Big) \bigg\} \, .
\end{eqnarray}
where $\hat{a},\hat{a}^{\dagger}$ stand for the ladder operators of the EM field. Non-conserving energy terms in Eq.~(\ref{eq:2})  are omitted. The first term in the Hamiltonian Eq.~(\ref{eq:1}) represents the photon \textit{up-conversion} assisted by graviton absorption, described by $\hat{a}_{k} \hat{a}^{\dagger}_{k'}  \hat{b}_{K}$ and happens when the energy condition $\omega_{k'} = \omega_{k} + \Omega_{K}$ is fulfilled. Photon \textit{down-conversion} assisted by graviton emission is described by $
\hat{a}_{k} \hat{a}^{\dagger}_{k'} \hat{b}^{\dagger}_{K}$ and happens when $ \omega_{k'}= \omega_{k} - \Omega_{K}$ is satisfied. The interaction rates of each of these processes are denoted by $\mathcal{G}_{\text{up}}$ and $\mathcal{G}_{\text{down}}$, respectively.
The last term in Eq.~(\ref{eq:1}) shows photon-graviton mixing: two-photon annihilation followed by single-graviton creation and vice versa. Basically, these processes illustrate the capability of spontaneous or stimulated emission and absorption of photons. These are described by $
\hat{a}_{k} \hat{a}_{k'} \hat{b}^{\dagger}_{K}$ and $ \hat{a}^{\dagger}_{k} \hat{a}^{\dagger}_{k'} \hat{b}_{K}$, and the interaction rate of these processes is shown by $\mathcal{G}_{\text{mix}}$, representing the photon-graviton mixing. In free space and in the absence of an external magnetic field or media, these processes occur when the exact energy-momentum conservation is fulfilled: $\mathbf{K} = \mathbf{k} + \mathbf{k}'$ and $\Omega_{K} = \omega_{k} + \omega_{k'}$. Since the EM modes and GWs are all on-shell, satisfying $k_{\mu}k^{\mu} = 0$, $k'_{\mu} k'^{\mu}=0$ and $K_{\mu} K^{\mu} =0$, the energy-momentum conservation $K_{\mu} = k_{\mu}+k'_{\mu}$ implies that in a 3WM in vacuum, the vectors $\mathbf{k}, \mathbf{k}'$ and $\mathbf{K}$ are necessarily co-aligned. However, the GW polarization tensor couples to the wave vector of the EM field. As in the GW detectors such as LIGO, the arm lengths of the interferometer detect spacetime oscillations caused by plus and cross polarizations since the effect on the frequency modulation is $\sim h_{ij}\, \hat{k}_i \hat{k}_j$. One ends up with the result that for a collinear propagation, the polarization selection rules, embarked in $\mathcal{G}_{\text{mix}}(k,k',K)$ forbid photon-graviton mixing. Hence, it is said that Lorentz invariance forbids the spontaneous conversion of photons into gravitons and vise versa. Practically, the translational invariance (which leads to momentum conservation $\delta^{(3)}(\mathbf{K} - \mathbf{k} - \mathbf{k}')$) is broken, for instance, by introducing an anisotropic media \cite{morgan1990phase, brandenberger2023graviton}, or adding external magnetic fields as Gertsenshtein proposed \cite{Gertsenshtein1962}. 

It has to be noted that the interaction Hamiltonian Eq.~(\ref{eq:1}) is the most general form of interaction between EM and GW fields in free space, up to linear order in perturbations $\mathcal{O}(h)$. In the adiabatic approximation where $\Omega_K\ll \omega_k, \omega_{k'}$, it recasts to the familiar intensity-dependent coupling already introduced in \cite{guerreiro2020quantum, arani2023sensing} and two-photon processes (proportional to $\hat{a}_k\hat{a}_{k'}$ and $\hat{a}^{\dagger}_k\hat{a}^{\dagger}_{k'}$) vanish, since their corresponding transition amplitude is sharply peaked around $\omega_k+\omega_{k'} = \Omega_K \simeq 0$. Thus, it is exactly the on-resonance condition that makes the photon-graviton exchange possible.


\section{\label{sec:3}Cavity-QED platform}

Alternatively, cavity-QED environment naturally provides such symmetry breaking through reflecting boundaries. The conducting boundaries confine the EM field into discrete standing wave eigen-modes, each of which is a coherent superposition of multiple plane wave components with different propagation directions. The boundary-induced mode quantization introduces an effective anisotropicity in the EM wave numbers along different spatial directions. For instance, in a rectangular cavity each transverse electric (TE) mode, labeled by $\alpha=(m,n,p)$, possesses distinct longitudinal and transverse components of the wavevector, thus defining a dispersion relation analogous to that in a nonlinear crystal, namely $\omega_{mnp} = c\sqrt{(\frac{m\pi}{L_x})^2 + (\frac{n\pi}{L_y})^2 + (\frac{p\pi}{L_z})^2}$. This intrinsic anisotropy supplies the necessary phase-matching structure for 3WM, effectively replacing the role of material anisotropy in conventional nonlinear optics. Consequently, there are specific polarization selection rules emerging from phase matching conditions between three modes $k_{\alpha}, k_{\beta}, K$. The coupling strength $\mathcal{G}_{\text{mix}}(k,k',K)$ thus takes non-vanishing values, depending on the specific configuration of GW and the EM modes contained in the cavity.

\begin{figure}[]
\centering
\includegraphics[
width=0.9\columnwidth]{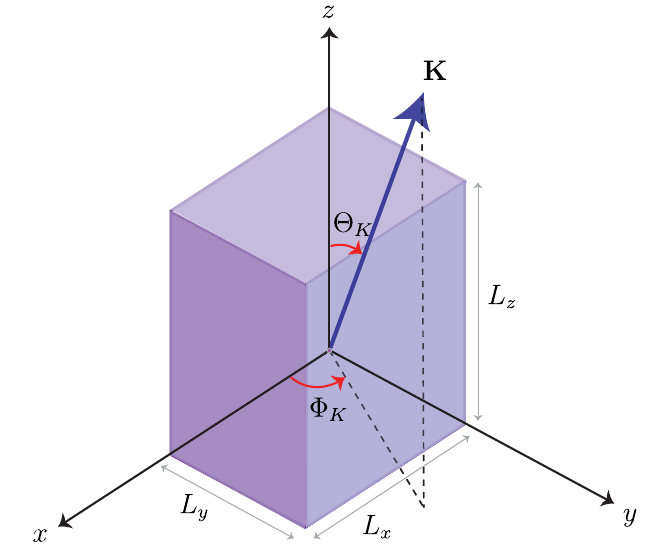}
\caption{Schematic view of a rectangular cavity of size $V=L_xL_yL_z$ illuminated by a plus-polarized mono-mode GW propagating in $\hat{\mathbf{K}}=(\sin\Theta_K\cos\Phi_K, \sin\Theta_K\sin\Phi_K,\cos\Theta_K)$ direction.}
\label{fig1}
\end{figure}

To see these clearly, we consider a rectangular cavity of size $V=L_x L_y L_z$  which is illuminated by a plus-polarized ($\gamma=+$) mono-mode GW which propagates in an arbitrary direction in space, represented by $\hat{\mathbf{K}}=(\sin\Theta_K\cos\Phi_K, \sin\Theta_K\sin\Phi_K,\cos\Theta_K)$. The schematic view is shown in Fig.~\ref{fig1}\,.
In order to find the eigen-modes of the cavity and to choose the Coulomb gauge, it is natural to adopt the Fermi normal modes, i.e., the proper detector frame (PDF), which describes the tidal gravitational field measured by the localized observer attached to the cavity. In the PDF, by solving Maxwell's equations and applying perfectly conducting boundary conditions and choosing the Coulomb gauge, the TE (as well as TM) modes are routinely calculable. 
Gravitational perturbations $h_{\mu\nu}$ must also be described in the PDF. It is then straightforward to transform the metric from TT gauge to the PDF (see \cite{ito2023exploring, navarro2024study}). 
%

Following the action method introduced in Sec.~\ref{sec:2}, one eventually ends up with a Hamiltonian of the form Eq.~(\ref{eq:1}) but where the coupling strength $\mathcal{G}_{\text{mix}}(k,k',K)$ is now symbolically replaced by $\mathcal{G}^{\gamma}_{\alpha\beta}(\mathbf{K})$ which shows the overlap coupling between two cavity TE modes $\alpha=(m,n,p)$ and $\beta=(m',n',p')$ and the incoming GW mode $K=(\mathbf{K}, \gamma)$. Note that the mixing process is resonant when the exact energy-match condition $\Omega_K=\omega_{\alpha} + \omega_{\beta}$ (or equivalently $|\mathbf{K}| = |\mathbf{k}_{\alpha}| + |\mathbf{k}_{\beta}|$) is met; otherwise, the transition probability of each process vanishes according to the Fermi's golden rule. 

In free space, mixing requires momentum conservation. In a cavity, however, translation invariance is broken, so the overlap integral does not yield delta functions. Instead, a finite-volume structure factor outcomes
\begin{eqnarray}\label{eq:3}
\mathcal{G}_{\alpha\beta}^{\gamma}(\mathbf{K}) &=& \sqrt{\omega_{\alpha}\omega_{\beta}}\, \mathcal{C}_g\, \mathcal{A}_{\alpha\beta}^{\gamma}(\mathbf{K}) \,,
\end{eqnarray}
where the overlap integral is defined by (see App.~\ref{app:A} for definition)
\begin{eqnarray}\label{eq:4}
\mathcal{A}_{\alpha\beta}^{\gamma}(\mathbf{K}) &\sim& \int_{V}d^3x \, (\text{EM-GW coupling tensors})\, e^{i\mathbf{K}\cdot\mathbf{x}}\,.\qquad
\end{eqnarray}
The integrand contains $\mathrm{sinc}(\Delta k_i L_i)$-like functions in three dimensions and phase factors of the form $e^{i\Delta k_i L_i}$, each peaked at small values of $\Delta k_i L_i \ll 1$ and decaying quickly with increasing it. Thus, the overlap integral is non-vanishing when the phase matching happens, i.e., when
\begin{eqnarray}\label{eq:5}
\Delta \mathbf{k} &\equiv& \mathbf{K} \pm \mathbf{k}_{\alpha} \pm \mathbf{k}_{\beta} \simeq 0 \, ,
\end{eqnarray}
where $\mathbf{k}_{\alpha}=(m,n,p)(\pi/L)$ and $\mathbf{k}_{\beta}=(m',n',p')(\pi/L)$ for a cubic cavity, and we assume $m,n,p,m',n',p' \geq 1$.
Hence, the overlap integral is the cavity version of momentum conservation;
it selects the GW wave vectors and directions that satisfy phase matching. It also encodes which GW polarizations couple to which cavity modes. It has to be noted that in the limit of a large cavity where $L\rightarrow \infty$, the $\mathrm{sinc}$ function tends to delta function which enforces the colinear propagation. In this manner, the overlap integral vanishes as one would expect for the free propagation of EM field. Of course this is a gauge-invariant consequence.
We also note that, in contrast to the Gertsenshtein mechanism, there is no external magnetic field $\mathbf{B}_0$ inside the cavity. Instead, the cavity mode itself provides a spatially varying internal magnetic field that mimics the role of an external magnetic field to mediate graviton-photon conversion.

In the following, we discuss some of the main results and implications of the trilinear Hamiltonian Eq.~(\ref{eq:2}). In order to solve the dynamics, it is helpful to find symmetries of the Hamiltonian. It is easy to show that the first two terms in the Hamiltonian Eq.~(\ref{eq:2}) conserve the total number of photons, while the last term, $\hat{a}^{\dagger}_{\alpha} \hat{a}^{\dagger}_{\beta}\hat{b}_{K}$, does not. 
In order to investigate photon creation by the GW field, it is hence sufficient to consider the last term. In other words, we can consider a situation where only two EM modes $\alpha=(m,n,p)$ and $\beta=(m',n',p')$ coexist and interact with a mono-mode plus-polarized GW.
Consequently, the two-mode photon creation by a single-mode GW is fully described by the trilinear Hamiltonian,
\begin{eqnarray} \label{eq:6}
\hspace*{-0.5cm}\hat{H} / \hbar &=& \omega_{\alpha} \hat{a}^{\dagger}_{\alpha}\hat{a}_{\alpha} + \omega_{\beta} \hat{a}^{\dagger}_{\beta}\hat{a}_{\beta} + \Omega_K \hat{b}^{\dagger}_{K} \hat{b}_{K} \\
&+&|g| \big( \hat{b}_{K} \hat{a}^{\dagger}_{\alpha} \hat{a}^{\dagger}_{\beta} +  \, \hat{b}^{\dagger}_{K} \hat{a}_{\alpha} \hat{a}_{\beta} \big) \,. \nonumber
\end{eqnarray}
In the above equation, the replacement $\mathcal{G}^{\gamma}_{\alpha\beta}(\mathbf{K}) \rightarrow g$ is used for abbreviation. Moreover, note that the interaction Hamiltonian Eq.~(\ref{eq:6}) is written in the form $\hat{H}_{int} = \hbar |g| \big( \hat{b}_K \hat{a}_{\alpha}^{\dagger}  \hat{a}_{\beta}^{\dagger} + \text{h.c.}\big)$ by defining $g \equiv |g| e^{i\phi}$, and one can can always perform a unitary phase rotation of one of the modes (e.g. $\hat{a}_{\alpha} \rightarrow e^{-i\phi} \hat{a}_{\alpha}$) to absorb the phase. 
The total number of particles is not conserved by the Hamiltonian Eq.~(\ref{eq:6}). However, if one defines the quantity $\hat{M} \equiv 2 \hat{n}_K + \hat{n}_{\alpha} + \hat{n}_{\beta}$ then it can be shown that $[\hat{H}, \hat{M}] = 0$, so that $\hat{M}$ is a conserved quantity, called as the Manley-Row invariant, which is a specific weighted combination of mode occupations.
The other conserved quantity is the difference photonic occupation number $\hat{N}_{\text{diff}} \equiv \hat{n}_{\alpha} - \hat{n}_{\beta}$ for which one can easily show that $\mathrm{d}\hat{N}_{\text{diff}}/\mathrm{d}t=0$. 
Hence, the system dynamics is restricted to Hilbert sub-spaces with definite $\langle \hat{M}\rangle$ and $\langle\hat{N}_{\text{diff}}\rangle$, specified by their initial values.


\section{\label{sec:4}Quantum aspects of GWs}

Since one is interested in photon creation from gravitons, the initial state of the cavity is taken as vacuum. The energy condition $\Omega_{K} = \omega_{\alpha} +\omega_{\beta}$ is the necessary condition of coupling. To see what happens, we assume that a GW of frequency $f\sim 3.9\,$GHz illuminates the cavity in a specific direction $\Theta_K = \pi/2, \Phi_K = \pi/6$. The resonance condition then implies that one of the modes is $\alpha=(1,1,1)$ while the other mode can be one of three modes $\vec{\beta}=\{(2,1,1),(1,2,1),(1,1,2)\}$: these modes are degenerate and possess identical energy $\omega_{211}=\omega_{121}=\omega_{112} \equiv \omega_{211}$. Numerical investigation shows that the overlap integral $\mathcal{A}^{+}_{111,\beta}$ is non-vanishing for these three combinations of $\alpha$ and $\beta$ and is given by $\vec{\mathcal{A}}^{+}_{111,\beta} \equiv \{ 1.75, 2.23, 1.44 \}$, where $\mathcal{A}^{+}_{111,\beta_i}$ shows the overlap integral between two modes $\alpha$ and $\beta_i$ for $i=1,2,3$.
As a result, the interaction term in the Hamiltonian Eq.~(\ref{eq:6}) contains $\sim \hat{b}_K \hat{a}_{\alpha} \big(\mathcal{A}_1 \hat{a}_{\beta_1} + \mathcal{A}_2 \hat{a}_{\beta_2} + \mathcal{A}_3 \hat{a}_{\beta_3} \big) + \text{h.c.}$. One can define a unitary transformation that transforms $\hat{a}_{\beta_i}$ to new collective modes $\hat{c}_{i}$. Since the old modes are degenerate, it follows that the collective mode $|\mathcal{A}|\, \hat{c}_1 \equiv \mathcal{A}_1 \hat{a}_{\beta_1} + \mathcal{A}_2 \hat{a}_{\beta_2} + \mathcal{A}_3 \hat{a}_{\beta_3}$
enters the interaction (called the bright mode), while $\hat{c}_2$ and $\hat{c}_3$ are irrelevant (called dark modes). Here, $|\mathcal{A}| \equiv \sqrt{\mathcal{A}_1^2 + \mathcal{A}_2^2 + \mathcal{A}_3^2}\simeq 3.18$. Consequently, the interaction reduces to that of the original trilinear form Eq.~(\ref{eq:6}), with $\hat{a}_{\beta}$ replaced by the collective mode $\hat{c}_1$. The dynamics of collective mode $\hat{c}_1$ then follows immediately.

Afterward, one drives $\hat{b}_{K}$ strongly by a coherent GW pump placed in a coherent state $|\eta_K\rangle $, with $n_g = |\eta|^2 \simeq M_{\text{pl}}^2 \frac{h_+^2}{f^2}$ graviton content (in natural units), where $h_+$ stands for the GW strain.
For typical values of $h_+\sim 10^{-21}$ and $f \sim 3.9\,$GHz one obtains $n_g\sim 10^{24}$, i.e, a huge number of gravitons. In the limit $\langle \hat{b}\rangle = \eta \gg 1$, the interaction between EM modes effectively reduces to a bilinear form $\hat{H}_{\text{int}} \propto |g\,\eta|\, (\hat{a}^{\dagger}_{\alpha} \hat{c}^{\dagger}_1 + \text{h.c.})\,$, e.g., a two-mode squeezer that entangles the EM modes. 
In this semiclassical approximation, we neglect quantum fluctuations of the GW field and consider it as a classical mean field. This approximation is valid as long as the mean number of gravitons in the field is much larger than the fluctuations of the field. Thus, the mean number of generated photons after time $t$ is $2\sinh^2 r$ with $r \equiv g\,\sqrt{n} \,t$ being the squeezing amplitude. Calculations show that the squeezing amplitude generated by a monochromatic coherent GW is
\begin{eqnarray}\label{eq:7}
r(t) &\sim& h_+ \,\Omega_K \, |\mathcal{A}|\, t  \, . 
\end{eqnarray}
In this manner, GW transforms into possibly detectable EM waves whose amplitude grows as $\exp \big(\epsilon h_+ \Omega_K\,t)$, where $\epsilon$ represents an order one constant that is determined from phase and polarization matching. This consequence is in accordance with the result of \cite{brandenberger2023graviton} where it is shown that due to the exponential instability induced by parametric resonance, the EM field undergoes amplification. In that setup, the necessity of a medium for the graviton-photon conversion is underlined, while here we obtained a similar result in a cavity platform.
For three-dimensional superconducting resonators working at $\sim $\,GHz frequency, maximum photon lifetimes up to $\sim 2$ seconds has been reported \cite{romanenko2020three}. Considering a typical value of the strain field $10^{-21}$ at $3.9$\,GHz frequency, one obtains $r\sim 10^{-12}$ that corresponds to the generation of $\bar{n}_{\text{ph}}^{(\text{sc})} \sim 3.8 \times 10^{-24}$ photons after $2$ seconds. Eq.~(\ref{eq:7}) implies that in order to detect squeezed photons with $r\sim 1$ converted from GW, one needs to wait for $\sim 10^{12}\,$s which is extremely large. 

On the other hand, if we stay at the quantum regime for the graviton mode, where both EM modes are initially in the vacuum state, photon pairs are generated purely through quantum fluctuations mediated by the graviton mode. The trilinear interaction induces transitions like 
\begin{eqnarray}\label{eq:8}
|n_g,0_{\alpha},0_{c}\rangle \rightarrow |n_g-1,1_{\alpha} , 1_{c}\rangle \rightarrow |n_g-2,2_{\alpha},2_{c}\rangle \cdot\cdot\cdot \, ,
\end{eqnarray}
within each invariant subspace of the Hilbert space, so that the Manley-Row invariant and the difference in photonic occupation number stay conserved.
As the graviton content $n_g$ increases, the characteristic timescale for photon production decreases as $t_{\text{sp}} \sim \frac{1}{g\sqrt{n_g}}$. Thus, increasing the initial graviton occupation accelerates spontaneous photon generation through collective bosonic enhancement. Fig.~\ref{fig3} shows the behavior of occupation numbers versus the reduced timescale $\tau=gt$. For the sake of illustration, the initial state of gravitons is taken as a coherent state with $n_g=50$ graviton content. The black curve shows the total EM excitation in the collective mode $\hat{c}$, starting from zero at the initial time. Generated photons in modes $\hat{a}_{\beta_1}, \hat{a}_{\beta_2}, \hat{a}_{\beta_3}$ are shown by red, blue and green curves. Graviton occupation is shown by the purple curve, starting from $n_g=50$. Semiclassical prediction of hyperbolic amplification is shown by the gray-dashed curve. Importantly, in the spontaneous quantum regime, the dynamics is bounded: photon number growth saturates due to graviton depletion, leading to oscillatory energy exchange rather than exponential amplification. This behavior is intrinsically quantum and reflects the finite-dimensional structure of the Hilbert space, imposed by the conserved Manley–Rowe invariant. The semiclassical and quantum behaviors are quite distinguished after a timescale $\tau= gt \sim \frac{1}{\sqrt{n_g}} \sim \frac{1}{5\sqrt{2}}$.
\begin{figure}[]
\centering
\includegraphics[
width=0.94\columnwidth]{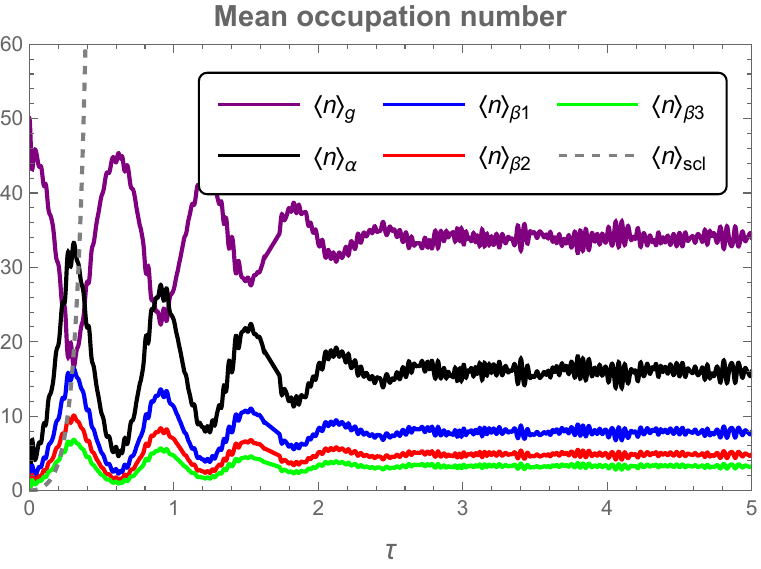}
\caption{The mean number of particles in GW mode $K$, collective EM mode $\hat{c}$ and EM degenerate modes $\hat{a}_{\beta_1}, \hat{a}_{\beta_2}, \hat{a}_{\beta_2}$ versus the dimensionless timescale $\tau = g t$, shown in purple, black, red, blue and green, respectively. Initial graviton content in coherent state is taken $n_g=50$. Note that the classical hyperbolic amplification is $2\sinh^2(r) = \sinh^2 (|g|\,\sqrt{24}\, t) = \sinh^2(\sqrt{24} \,\tau)$, as shown by the gray dashed line.}
\label{fig2}
\end{figure}

Physically, spontaneous emission corresponds to graviton-induced pair creation from vacuum fluctuations of the cavity field.
Each conversion event entangles the gravitational and EM sectors, which will cause entanglement degradation between $\hat{a}_{\alpha}$ and $\hat{c}$ modes. Hence, the EM reduced density matrix $\hat{\rho}_{\alpha c} = \text{Tr}_K[|\psi(t)\rangle \langle \psi(t)|]$ becomes mixed. This implies that the gravitational quantum pump is not just a background but a quantum part of the total system. The purity of the EM field is defined as $ \mu_{\alpha c}(t) = \text{Tr}_{\alpha c}[\hat{\rho}_{\alpha c}^2(t)]$ which takes values $\in [0,1]$.
At short times, the entanglement generation rate scales with $n_g$. In Fig.~\ref{fig3} the EM purity versus the reduced timescale $\tau=gt$ is shown. It shows a rapid decrease from the reference value of unity and eventually fluctuates around a small value. The gray-dashed line refers to the purity of the two-mode squeezed photons predicted in the semiclassical regime, which is a pure state with the highest purity. Increasing graviton occupation accelerates entanglement generation between EM and GW sectors, causing the EM subsystem to become mixed more rapidly.
\begin{figure}[]
\centering
\includegraphics[
width=0.99\columnwidth]{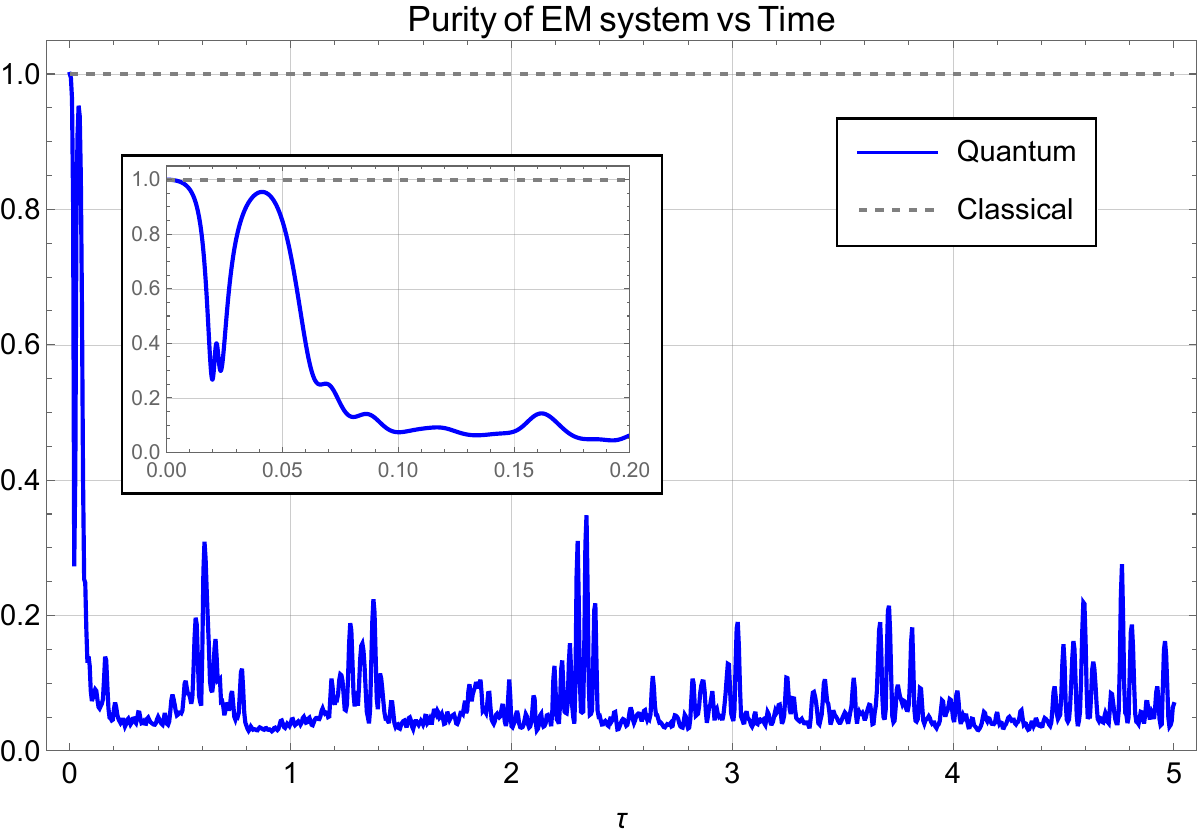}
\caption{The purity $\mu_{\alpha c}(\tau)$ versus the reduced timescale $\tau=gt$. The reference purity (classical GW) $\mu_{\alpha c}=1$ is shown by the gray dashed line. The inset shows the early-time abrupt decrease of purity.}
\label{fig3}
\end{figure}

Usually, energy-based observables, such as single-graviton Rabi oscillations or saturation, require substantial photon number transfer and therefore long interaction times. Detecting full energy exchange between gravitational and EM modes is extremely challenging given the smallness of $g$. In contrast, purity decay is sensitive to the buildup of quantum correlations rather than to macroscopic energy transfer. Even when photon occupation remains small, coherent graviton–photon entanglement can already be significant.
Experimentally, electromagnetic purity can be accessed via quantum state tomography, homodyne detection, or quadrature measurement techniques that are well developed in cavity and circuit QED. Thus, entanglement-based observables provide a potentially more sensitive probe of graviton quantization than direct detection of energy exchange.

At last, stimulated emission provides an additional acceleration mechanism beyond graviton occupation alone. The presence of initial photons in either mode enhances the transition amplitudes along the pair-creation ladder, leading to faster photon growth in the idler mode and more rapid entanglement generation. More precisely, if the system starts from an initial state $|\psi(0)\rangle = |n_g, n_{\alpha},0\rangle$, so that the difference photon number is $n_{\alpha}$, then using a Holstein–Primakoff representation of the SU(1,1) generators, the interaction Hamiltonian reduces at early times to an effective beam-splitter coupling between the graviton mode and a collective photon-pair mode, $\hat{H}_{int} \propto \hbar g \sqrt{n_{\alpha}+1}\,(\hat{b}_K \hat{C}^{\dagger}+\text{h.c.})$, where $\hat{C}$ represents the collective EM mode excitations including $\alpha$ and $\beta_{i}$ modes. Hence, the characteristic timescale scales as $t_{\text{stim}}\sim\frac{1}{g\sqrt{n_g (n_{\alpha}+1)}}$. Thus, stimulated emission provides an additional acceleration mechanism beyond graviton occupation alone. The time-scaling $t_{\text{sim}} \propto \frac{1}{n^2}$ is reminiscent of the Dicke-type superradiance \cite{dicke1954coherence}, as this analogy between Dicke model and trilnear Hamiltonian has already been considered in the literature, in the context of nonlinear optics \cite{walls1970quantum} and trilinear Hamiltonian modeling of the Hawking radiation \cite{bambah2013entanglement}. Here, we generalize the analogy to the resonant EM-GW system. Hence, the presence of initial photons enhances the transition amplitudes, leading to faster photon growth and more rapid entanglement generation. Taking the previous value $g\sqrt{n}_g \sim 10^{-12}$, if one starts from a highly occupied EM mode with $n_{\alpha} \sim 10^{24}$, the effective coupling scales as $\sim \mathcal{O}(1)$. Subsequently, this approach may provide a promising route to circumvent astronomically long observation times and bring the required timescales into a regime that is experimentally plausible in the laboratory.


\section{\label{sec:5}Conclusion}

We have analyzed photon generation mediated by a quantized gravitational wave mode inside a cavity-QED platform. This resonant interaction is governed by a trilinear Hamiltonian, which enables coherent conversion between one graviton and a correlated photon pair, reducing the EM–GW coupling to an exactly solvable three-wave mixing quantum problem. The present platform makes it possible to explore classical and quantum characteristics of photon-graviton mixing and to distinguish between different regimes.
In contrast to semiclassical hyperbolic amplification of photons, the fully quantum treatment predicts saturated, oscillatory energy exchange due to pump depletion. The dynamics remain confined within invariant subspaces determined by conserved Manley–Rowe quantities, replacing exponential growth with saturation.
Beyond photon-number dynamics, graviton–photon interaction generates entanglement between gravitational and electromagnetic modes. 
The characteristic time scale scales as $t_{\text{sp}}\sim (g\sqrt{n_g})^{-1}$, demonstrating collective bosonic enhancement of the interaction. In the stimulated regime, where one electromagnetic mode is initially populated with $n_{\alpha}$ quanta in either modes, the effective coupling is further enhanced, leading to $t_{\text{stim}}\sim (g\sqrt{n_g(n_{\alpha}+1)})^{-1}$. Thus, both gravitational and electromagnetic occupations reduce the interaction time through square-root collective scaling, analogous to Dicke-type superradiant emission.
Altogether, the trilinear Hamiltonian provides a transparent and quantitative framework for comparing semiclassical and fully quantum predictions, and for identifying the qualitative features that arise uniquely in the quantum regime. It also offers a conceptually clean setting in which photon-graviton three-wave mixing can be studied without relying on large external magnetic fields.
Although the fundamental interaction remains Planck suppressed, collective occupation of the gravitational and electromagnetic modes provide a clear mechanism for reducing the relevant dynamical time scales. The cavity-QED framework therefore recasts electromagnetic-gravitational coupling as a controllable quantum-optical process and furnishes a useful platform for investigating collective enhancement, pump depletion, and entanglement-based signatures of quantum gravitational mediation.

\begin{acknowledgements}
J.\ S. was in part supported by JSPS KAKENHI Grant Numbers JP23K22491, JP24K21548, JP25H02186.
\end{acknowledgements}

\bibliography{biblio}


\section{Supplementary material} \label{sec:6}

\appendix

\section{The overlap integral $\mathcal{A}_{\alpha,\beta}^{\lambda}(\mathbf{K})$} \label{app:A}

Assuming that the electromagnetic TE modes inside the cavity are represented by eigenmodes $\mathbf{u}_{\alpha}(\mathbf{x})$ and performing the the canonical quantization of the Hamiltonian in the proper detector frame, the expression of the overlap integral is given by
\begin{widetext}
\begin{eqnarray}\label{eq:appA1}
\mathcal{A}_{\alpha\beta}^{\gamma}(\mathbf{K}) &\equiv& \int_{cavity} dV\, \bigg\{ \frac{1}{4}\, \Xi_{ijkl}(\mathbf{x})\, \Big( \frac{1}{k_{\alpha}k_{\beta}} (\nabla\times \mathbf{u}_{\alpha})_i\, (\nabla\times \mathbf{u}_{\beta})_j - u_{\alpha i} \, u_{\beta j} \Big) - \frac{i}{2\,k_{\beta}} \,\epsilon_{ijn} \, u_{\alpha i} \, (\nabla \times \mathbf{u}_{\beta})_j\, \Gamma_{nkl}\,  \bigg\}\, x^k x^l \, . \nonumber\\
\end{eqnarray} 
\end{widetext}
Here, GW tensors $\Xi_{ijkl}(\mathbf{x})$ and $\Gamma_{nkl}$ are defined by
\begin{widetext}
\begin{eqnarray}\label{eq:appA2}
\Xi_{ijkl}(\mathbf{x}) &\equiv& - \Big( K_j K_k e_{il}^{\lambda} + K_l K_i e_{kj}^{\lambda} - K_l K_k e_{ij}^{\lambda} - K_i K_j e_{kl}^{\lambda} \Big) \, F_2(\mathbf{K}\!\cdot\!\mathbf{x}) + \frac{K^2}{2}\,\delta_{ij}\, e^{\lambda}_{kl}\, \big( F_0(\mathbf{K}\!\cdot\!\mathbf{x}) - F_2(\mathbf{K}\!\cdot\!\mathbf{x})\big)\, ,\nonumber\\
\Gamma_{nkl} (\mathbf{x}) &\equiv& K \, \big( K_n e_{kl} - K_l e_{nk} \big) \, F_1(\mathbf{K} \cdot \mathbf{x}) \,, \nonumber
\end{eqnarray} 
\end{widetext}
where 
 \begin{eqnarray}\label{eq:appA3}
F_0(u) &=& \Re\Big[ \frac{1-e^{-iu}}{u^2} - \frac{i}{u} \Big] = \frac{1-\cos u}{u^2}\,, \\
F_1(u) &=& \Re\Big[ i \frac{1-e^{-iu}}{u^3} + \frac{i}{2u} + \frac{e^{-iu}}{u^2} \Big] = \frac{-\sin u}{u^3} + \frac{\cos u}{u^2} \, , \nonumber\\
F_2(u) &=& \Re\Big[ \frac{1+e^{-iu}}{u^2} + 2i \frac{1-e^{-iu}}{u^3} \Big] = \frac{1+\cos u}{u^2} - \frac{2 \sin u}{u^3} \, . \nonumber
\end{eqnarray}

\end{document}